\documentclass{article}
\usepackage{dcase2018,amsmath,graphicx,url,times,booktabs, tabularx}

\title{Convolutional Neural Networks and x-vector Embedding for DCASE2018 Acoustic Scene Classification Challenge}

\name{Hossein Zeinali, Luk\'{a}\v{s} Burget and Jan ``Honza'' \v{C}ernock\'{y}}
\address{Brno University of Technology, Speech@FIT and IT4I Center of Excellence, Czech Republic}

\begin{document}

\ninept
\maketitle

\begin{sloppy}

\begin{abstract}
In this paper, the Brno University of Technology (BUT) team submissions for Task 1 (Acoustic Scene Classification, ASC) of the DCASE-2018 challenge are described. Also, the analysis of different methods on the leaderboard set is provided. The proposed approach is a fusion of two different Convolutional Neural Network (CNN) topologies. The first one is the common two-dimensional CNNs which is mainly used in image classification. The second one is a one-dimensional CNN for extracting fixed-length audio segment embeddings, so called x-vectors, which has also been used in speech processing, especially for speaker recognition. In addition to the different topologies, two types of features were tested: log mel-spectrogram and CQT features. Finally, the outputs of different systems are fused using a simple output averaging in the best performing system. Our submissions ranked third among 24 teams in the ASC sub-task A (task1a).
\end{abstract}

\begin{keywords}
Audio scene classification, Convolutional neural networks, Deep learning, x-vectors, Regularized LDA
\end{keywords}

\section{Introduction}
\label{sec:intro}

This paper deals with the problem of classifying scene or environment (see examples listed in Table~\ref{tbl.fold_results}) based on acoustic clues, which are normally used by humans and animals to understand and react on different environmental condition.
%ASC is one of the interesting research areas of non-speech audio processing and the community is tankful to the DCASE organizers for managing the annual challenges, which allowed for rapid advances in the ASC technology.
Several methods have been proposed for the Acoustic Scene Classification (ASC). Nowadays, most of them are deep learning based. The winner of the last year ASC challenge (i.e. DCASE2017 Task1) used Generative Adversarial Network (GAN) for data augmentation and the combination of Support Vector Machine (SVM) and CNN for classification~\cite{mun2017}. The most used network topology in the previous challenges is CNN proven to provide very good performance for ASC~\cite{mun2017,han2017,weiping2017,hyder2017}. The winner of DCASE2016 Challenge Task1~\cite{eghbal2016cp} also used CNN fused with an i-vector based method~\cite{dehak2011front}.

This report describes Brno University of Technology (BUT) team submissions for the ASC challenge of DCASE 2018. We proposed two different deep neural network topologies for this task. The first one is a common two-dimensional CNN network for processing audio segments as fixed size two-dimensional images. This network is fed in two ways: with single channel features and 4-channels features. This type of CNN network is useful for detection of audio events invariant to their position in audio signals. The second network topology uses a one-dimensional CNN along the time axis and is used to extract fixed-length embeddings of (possibly variable length) acoustic segments. This architecture has been previously found useful for other speech processing tasks such as speaker recognition~\cite{snyder2016deep}, where the extracted embeddings were called x-vectors. Therefore, in the rest of the paper, we will also refer to such neural embeddings of acoustic segments as to {\em x-vectors}. These networks were trained with two feature types: log mel-spectrogram and constant-Q transform (CQT) features. Our submissions are based on fusions of different networks and features trained on the original development data or using additional augmented data.

The current ASC challenge has three sub-tasks: In task1a, participants are allowed to use only the fixed development data for training. Task1b is similar to task1a except that the test files are from different mobile channels. Finally, task1c evaluation data is the same as task1a but additional data is allowed for training. We have participated in task1a only.

\section{Dataset}

In this work, the DCASE2018 data was used~\cite{mesaros2018multi}. The dataset consists of recordings from 10 scene classes and was acquired in six large European cities, in different environments in each city. The development set of the dataset consists of 864 segments for each acoustic scene which means a total of 8640 audio segments. The evaluation set was collected in the same cities, but in different environments and has 3600 audio segments. Each segment has an exactly 10-second duration, this is achieved by splitting longer audio recordings from each environment. The dataset includes a predefined validation fold. Each team can also create its own folds, but we used the single official fold for evaluation. The audio segments are 2-channels stereo files, recorded at 48\:KHz sampling rate.

\section{Data Processing}

\subsection{Features}

In this work, different features are used in single and multichannel modes. All features are extracted from zero mean audio signals. The main features are log mel-scale spectrogram. For extracting these features, first short time Fourier transform is computed on 40 ms Hamming windowed frames with 20 ms overlap using 2048 point FFT. Next, the power spectrum is transformed to 80 Mel-scale band energies and, finally, log of these energies is taken. The second set of features is obtained as 80-dimensional constant-Q transform of audio signals~\cite{schorkhuber2010constant}. This features are extracted using librosa toolbox~\cite{mcfee2015librosa}.

We used the features in two modes, single-channel and 4-channels. In single channel mode, the audio signal is first converted to mono and single-channel features are extracted from it (these features are indicated by ``M'' in the tables). In the 4-channels mode, four sets of features are extracted from the signal similar to~\cite{han2017} (these features are indicated by ``LRMS'' in the tables). Two feature sets from left (L) and right (R) channels, one from the summation of both channels (i.e. $M = L + R$) and one from the subtraction of both channels (i.e. $S = L - R$). We use these 4 feature sets as a single input to the CNNs. This mode is similar to multi-channel images (e.g. RGB channels), which are the typical CNN inputs in image classification. In previous works~\cite{han2017, eghbal2016cp}, each channel was processed separately and final scores were obtained by fusion of different channel scores. Here, the network tries to use all channels at the same time to use all the available information.

\subsection{Data augmentation}

Different methods have been proposed for data augmentation in audio processing. Based on the rules of the challenge task1a, external data cannot be used for the data augmentation. Because of this limitation and based on our initial experiments, we decided to use a simple method based on the assumption that a combination of two or more audio segments from the same scene is another sample of that scene with more complex pattern and events. Two new segments were generated for each audio segment as a weighted sum of the audio and several other randomly selected audios from the same scene. This way, we have tripled the amount of training data.

\section{CNN topologies}

We have used two different CNN topologies for this challenge. The first one is the common two-dimensional CNN known from image processing and the second topology is a one-dimensional CNN for extracting x-vectors -- neural network embeddings of audio segment as used, for example, in speaker recognition~\cite{snyder2016deep}. Both networks are described in more detail in the following sections.

\subsection{Two-Dimensional CNN}

We followed the common CNN framework proposed in~\cite{hyder2017} with some modifications. Table~\ref{tbl.cnn_topo} shows the network architecture. The network contains 3 CNN blocks. The first layer is a two-dimensional convolutional layer with 32 filters with kernel size $7 \times 11$ and unitary depth and stride in both dimensions. This layer is followed by batch-normalization and {\em  Rectified Linear Unit (ReLU)} activations. The next layer is a max-pooling layer operating over $2 \times 10$ non-overlapping rectangles, which is followed by the dropout layer at the end of the CNN block. The output of this block form the input to the next block and so on. The filter and kernel sizes of each layer are shown in Table~\ref{tbl.cnn_topo}. The last MaxPooling layer in the network operates over the entire time sequence length (i.e. the output of the layer has dimension one for the time axis). The next layer after the third CNN block is a global average pooling (over the frequency axis), which is followed by a batch-normalization layer. Finally, the last layer of the network is a Dense layer (fully connected) with 10 nodes and the softmax activation function.
Compared to~\cite{hyder2017}, where only one-channel features were used as the CNN input, we also train another CNN with 4-channel features (as indicated in the first line of Table~\ref{tbl.cnn_topo}).

\begin{table}[t]
  \renewcommand{\arraystretch}{1.0}
  \caption{\label{tbl.cnn_topo} 2-Dimensional CNN topology. BN: Batch Normalization, ReLU: Rectifier Linear Unit. The numbers in the parentheses show the kernel size of convolution layer and the number before BN shows the filter size of the layer. The numbers before MaxPooling show the window size for this layer.}
  \vspace{2mm}
  \centering{
    \setlength\tabcolsep{8pt}
    \begin{tabular}{ c }
    Input $80 \times 500 \times 1$ or $80 \times 500 \times 4$\\
    \midrule
    ($7 \times 11$) Conv2D(pad=1, stride=1)-32-BN-ReLU \\
    ($2 \times 10$) MaxPooling2D \\
    Dropout (0.3) \\
    \midrule
    ($7 \times 11$) Conv2D(pad=1, stride=1)-64-BN-ReLU \\
    ($2 \times 5$) MaxPooling2D \\
    Dropout (0.3) \\
    \midrule
    ($7 \times 11$) Conv2D(pad=1, stride=1)-128-BN-ReLU \\
    ($5 \times 10$) MaxPooling2D \\
    Dropout (0.3) \\
    \midrule
    GlobalAveragePooling2D \\
    BatchNormalization \\
    \midrule
    Dense-10-SoftMax \\
    \end{tabular}
    \vspace{-5mm}
  }
\end{table}

\subsection{One-dimensional CNN for x-vector extraction}

The CNNs extracting x-vectors use one-dimensional convolution along the time. Table~\ref{tbl.xvector_topo} shows the network architecture. The network has three parts. The first part operates on the frame-by-frame level and outputs sequence of activation vectors (one for each frame). The second part compresses the frame-by-frame information into a fixed length vector of statistics describing the whole acoustic segment. More precisely, mean and standard deviation of the input activation vectors are calculated over frames. The last part of the network consists of two Dense ReLU layers followed by a Dense softmax layer like in the previous topology. This network has been used in two ways: In the first case, the softmax output is used as before directly for the classification (i.e. we train end-to-end ASC system). In the second case, the x-vectors extracted at the output of the first affine transform after the pooling are used as the input for another classifier.

Linear Discriminant Analysis (LDA) transformation is used to precondition the x-vectors for the following ASC classifier (i.e. Cosine similarity classifier). More specifically, it is used to whiten the within-class covariance and possibly reduce the dimensionality of the x-vectors. For this purpose, the conventional LDA can be used, however, the number of preserved dimensions is at most the number of classes minus one (9 in our case). Our previous works in the text-dependent speaker verification~\cite{zeinali2016deep, zeinali2017hmm} and also the ASC experiments here indicate that such dimensionality reduction impacts the performance. For overcoming this limitation, we have proposed to use Regularized version of LDA (RLDA), which enables us to keep as many dimensions as we need. In RLDA, a small fraction of Identity matrix is added to both within and between-class covariance matrices giving the following estimation formulas:
\begin{eqnarray*}
    \mathbf{S}_w & = & \alpha\mathrm{\mathbf{I}} + \frac{1}{C}\sum_{c=1}^{C}\frac{1}{N_c}\sum_{n=1}^{N_c}
    (\mathbf{w}_c^n-\overline{\mathbf{w}}_c)(\mathbf{w}_c^n-\overline{\mathbf{w}}_c)^T\:, \\
    \mathbf{S}_b & = & \beta\mathrm{\mathbf{I}} + \frac{1}{C}\sum_{c=1}^{C}(\overline{\mathbf{w}}_c - \overline{\mathbf{w}})(\overline{\mathbf{w}}_c - \overline{\mathbf{w}})^T\:,
\end{eqnarray*}
where $\mathrm{\mathbf{I}}$ is the identity matrix, $C$ is the total number of classes (i.e. scenes in this case), $N_c$ is the number of training samples in class $c$, $\mathbf{w}_c^n$ is the $n^{\mathrm{th}}$ sample in class $c$, $\overline{\mathbf{w}}_c=\frac{1}{N_c}\sum_{n=1}^{N_c}\mathbf{w}_c^n$ is the mean of class $c$, $\overline{\mathbf{w}}=\frac{1}{C}\sum_{n=1}^{C}\overline{\mathbf{w}}_c$ is the mean of the class means and $\alpha$ and $\beta$ were empirically set to 0.001 and 0.01, respectively. This type of regularization makes the between-class covariance matrix of full rank, which allows us to freely choose the number of dimensions that we wish to preserve after the LDA transformation.
%\footnote{Out of the dimensions that would be discarded by the conventional LDA because of the zero between-class variability, we now preserve additional dimensions with the lowest within-class variability}
In this work, we reduce the original 128-dimensional x-vectors to 100 dimensions. For more information about RLDA, we refer readers to our previous papers~\cite{zeinali2017hmm,zeinali2017sut}.

After applying RLDA, average class x-vectors are estimated on training data and used as class representation vectors. Cosine similarity is calculated between each test x-vector and each class representation vectors and the class with the highest score is selected. Alternatively, these similarity scores are fused with other scores from the CNN outputs for the final decision making.

\begin{table}[t]
  \renewcommand{\arraystretch}{1.0}
  \caption{\label{tbl.xvector_topo} 1-Dimensional CNN topology for x-vector extraction. BN: Batch Normalization, ReLU:Rectifier Linear Unit.}
  \vspace{2mm}
  \centering{
    \setlength\tabcolsep{8pt}
    \begin{tabular}{ c }
    Input $500 \times 80$\\
    \midrule
    ($3 \times 1$) Conv1D(pad=1, stride=1)-128-ReLU-BN \\
    Dropout (0.15) \\
    \midrule
    ($3 \times 1$) Conv1D(pad=1, stride=1)-128-ReLU-BN \\
    Dropout (0.15) \\
    \midrule
    ($5 \times 1$) Conv1D(pad=1, stride=1)-128-ReLU-BN \\
    Dropout (0.15) \\
    \midrule
    ($1 \times 1$) Conv1D(pad=1, stride=1)-128-ReLU-BN \\
    Dropout (0.15) \\
    \midrule
    ($1 \times 1$) Conv1D(pad=1, stride=1)-256-ReLU-BN \\
    \midrule
    Statistic Pooling, Mean and Standard-Deviation \\
    \midrule
    Dense-128-ReLU-BN (x-vector)\\
    Dropout (0.15) \\
    \midrule
    Dense-128-ReLU-BN \\
    \midrule
    Dense-10-SoftMax \\
    \end{tabular}
    \vspace{-5mm}
  }
\end{table}

\section{Systems and Fusion}

In this challenge, we fused outputs of different systems to obtain the final results. For two-dimensional CNNs, both the single-channel and the 4-channels variants are trained on both sets of features, which gives us 4 different classifiers. Further, two CNN for x-vector extraction are trained each on one set of features. These are trained only for the single-channel variant. The softmax outputs of all 6 neural networks are directly used for classification. The two sets of x-vectors produced by the two latter CNNs are further used to construct another two cosine similarity based classifiers.

We trained these systems in two scenarios, the first one using the data without any augmentation and the second one using augmented data. The scores from the resulting 16 systems (8 for each scenario) were fused to form the final submission. We used two different strategies for system fusion: Multiclass logistic regression classifier was trained on the scores from the different systems outputs. FoCal Multiclass toolbox~\cite{brummer2007focal} was used for the logistic regression training. As an alternative fusion approach, we simply averaged the scores from the different systems. We used this alternative fusion strategies as we feared that the data available for the logistic regression fusion training might not be sufficient. The logistic regression classifier was trained on the validation set, which was already used for the early-stopping of CNN training and for the model selection (i.e. models performing best on the validation set were selected). Also, this set is rather small, which might lead to over-fitting during the fusion training.

The four final submissions to the challenge were system fusions obtained with the two fusion methods. Each method was used to fuse either 1) all the sub-systems trained only on the augmented data or 2) all the subsystems (i.e. also including the subsystems trained only on the original data).

\section{Experimental Setups}

The experiments reported in this section were mainly carried out on the official challenge validation fold, which divides the development set into two subsets: {\em training-set} and {\em evaluation-set}. There are 6122 and 2518 audio segments in each subset, respectively. The training set was further randomly divided into two separate parts with the portions of 70 and 30 percent. The bigger part of the training set was used for network training as well as classifier training (for the cosine distance based method), the smaller part of this set was used for stopping criteria in networks training, model selection and also the fusion training. Finally, the evaluation part was used for reporting the results.

In addition to the results on the development set, some results are reported using Kaggle leaderboard system\footnote{https://www.kaggle.com/c/dcase2018-task1a-leaderboard} on a leaderboard set, which has 1200 segments. This set was divided to public and private leaderboard subsets by the organizers and we report the results only for the public subset\footnote{The public subset has the same number of audio segments for each class and also is included in the evaluation set. As organizers mentioned, the results on the private subset are not valid because there are different numbers of audio segments per class.}. In this case, the whole development set was used for training and validation: about 90\% randomly selected audio segments of this set were used for training and other segments were used for validation. For the final system training, the same data split was used as for the leaderboard results. The final decisions for 3600 evaluation audio segments was submitted to the challenge website. For final submitted systems, the results on the evaluation set are also reported.

Similar to the baseline system provided by the organizers, our networks training was performed by optimizing the categorical cross-entropy using Adam optimizer~\cite{kingma2014adam}. The initial learning rate was set to 0.001 and the network training was early-stopped if the validation loss did not decrease for more than 20 epochs. Then, the training was started again from the best model but now with a reduced learning rate (half value). This training procedure is repeated 3 times until the learning rate reaches 0.00025. The maximum number of epochs and the mini-batch size were set to 200 and 64, respectively.

%The network topologies and their parameters are shown in Tables~\ref{tbl.cnn_topo} and~\ref{tbl.xvector_topo}. For the CNNs extracting embddings, we only used single-channel features for this challenge, but for 2-dimensional CNN both single and 4-channel features were fed to the different networks.

\section{Results}

\subsection{Comparison of Results}

Table~\ref{tbl.comarison_results} reports the public leaderboard results for individual systems as well as several system combinations. We separately report results for the systems using the two different feature sets in order to compare their performance. For the four systems submitted to the challenge, the table also provides the results on the evaluation set.

\begin{table}[t]
  \renewcommand{\arraystretch}{1.0}
  \caption{\label{tbl.comarison_results} Comparison results between different methods and feature types as well as two different fusion strategies and using data-augmentation or not. The star-marks on some fusion systems highlight the systems which were submitted as four final submissions to the challenge. M: single channel feature, LRMS: 4-channels feature, COS: cosine distance, MEL-All: all systems with MEL features and similarly for CQT-all.}
  \vspace{2mm}
  \centerline
  {
  \setlength\tabcolsep{4pt}
    \begin{tabular}{ l c c c }
    \toprule
    \midrule
    Method & & Public ACC [\%] & Eval. ACC [\%] \\
    \midrule
    Baseline system		& & 62.5 & 61.0 \\
    \midrule
    \multicolumn{4}{c}{Without data augmentation} \\
    \midrule
    Mel-2D-CNN-M		& & 71.0 &  \\
    Mel-2D-CNN-LRMS		& & 67.7 &  \\
    Mel-1D-CNN  		& & 65.3 &  \\
    Mel-x-vector-cos	& & 64.8 &  \\
    CQT-2D-CNN-M		& & 67.8 &  \\
    CQT-2D-CNN-LRMS		& & 68.8 &  \\
    CQT-1D-CNN  		& & 60.3 &  \\
    CQT-x-vector-cos	& & 60.2 &  \\
    Fusion-Average		& & \bf{75.0} &  \\
    Fusion-FoCal		& & 71.5 &  \\
    \midrule
    \multicolumn{4}{c}{With data augmentation} \\
    \midrule
    Mel-2D-CNN-M		& & 68.2 &  \\
    Mel-2D-CNN-LRMS		& & 71.3 &  \\
    Mel-1D-CNN		& & 67.8 &  \\
    Mel-x-vector-cos	& & 64.7 &  \\
    CQT-2D-CNN-M		& & 64.8 &  \\
    CQT-2D-CNN-LRMS		& & 68.5 &  \\
    CQT-1D-CNN		& & 60.8 &  \\
    CQT-x-vector-cos	& & 58.2 &  \\
    Fusion-Average $^*$	& & \bf{76.8} & \bf{78.1} \\
    Fusion-FoCal $^*$	& & 73.3 & 75.1 \\
    \midrule
    \multicolumn{4}{c}{Fusions} \\
    \midrule
    MEL-All-Average		& & 72.5 &  \\
    CQT-All-Average		& & 71.3 &  \\
    All-Average $^*$	& & \bf{77.5} & \bf{78.4} \\
    All-FoCal $^*$		& & 73.0 & 74.5 \\
    \midrule
    \bottomrule
    \end{tabular}
    \vspace{-4mm}
  }
\end{table}

Comparing the results of the different features, we can see that the mel-spectrogram performs better for ASC task in all cases. However, the fusion of both feature sets improves the performance considerably, which indicates their complementarity.

Generally, feeding the networks with 4-channels features improves the performance as compared to the single-channel variant, especially when more training data is available by the data augmentation. In some cases, this strategy, however, degrades the performance. We believe it should generally improve it, so these cases deserve a further investigation.

When comparing the results from the first and the second sections of Table~\ref{tbl.comarison_results}, it is obvious that the augmentation helps in some situations but degrades the performance of other ones. The results are not consistent for all network types. As mentioned before, in four-channel modes the augmentation improves the performance in almost all cases.

The results from the two different fusion strategies show that the simple averaging performs considerably better in all cases. As we expected, the data for the fusion training were not sufficient. The fusion training over-fitted to the validation data and did not generalize well on other datasets.

The results on both leaderboard and evaluation sets show that the fusion of the 8-systems trained on the augmented data already achieves very good performance. When the systems with no data augmentation are also added to the fusion, only slight improvement can be obtained.

\subsection{Results on the Official Fold}

In this section, the results of the best final system (i.e. All-Average system from Table~\ref{tbl.comarison_results}) for each scene are reported. Table~\ref{tbl.fold_results} shows the performance of the system for each scene separately as well as the overall performance on the official challenge validation fold. The results indicate that our systems perform well for all the scene classes except the {\em Public Square} class, which deserves a future investigation.

\begin{table}[t]
  \renewcommand{\arraystretch}{1.0}
  \caption{\label{tbl.fold_results} Comparison results between different scenes of the final fused system.}
  \vspace{2mm}
  \centerline
  {
  \setlength\tabcolsep{8pt}
    \begin{tabular}{ l c c c c }
    \toprule
    \midrule
            			& & Our system      & Baseline \\
    Scene label			& & Accuracy [\%]   & Accuracy [\%]\\
    \midrule
    Airport				& & 91.6 &          72.9 \\
    Bus					& & 71.0 &          62.9 \\
    Metro				& & 78.4 &          51.2 \\
    Metro Station		& & 79.2 &          55.4 \\
    Park				& & 88.4 &          79.1 \\
    Public Square		& & 29.9 &          40.4 \\
    Shopping Mall		& & 77.5 &          49.6 \\
    Street Pedestrian	& & 75.4 &          50.0 \\
    Street Traffic		& & 82.0 &          80.5 \\
    Tram				& & 80.1 &          55.1 \\
    \midrule
    Average				& & 75.3 &          59.7 \\
    \midrule
    \bottomrule
    \end{tabular}
    \vspace{-4mm}
  }
\end{table}

\section{Conclusions}

We have described the systems submitted by BUT team to Acoustic Scene Classification (ASC) challenge of DCASE2018. Different systems were designed for this challenge and the final systems were fusions of the output scores from the individual system. A simple score averaging and logistic regression were used for the fusion. The systems included 2-dimensional CNNs with single and 4-channels features, one-dimensional CNNs trained on mel-spectrogram and CQT features. Cosine similarity classifiers were also used to compare x-vectors extracted using the one-dimensional CNNs.

Our future work will include investigations into the failures of the 4-channel CNN variants in some scenarios. We will also experiment with other methods for data augmentation, which, in our opinion, is crucial for the good system performance. Also, we would like to investigate into using bottleneck features for ASC.

\section{Acknowledgment}

The work was supported by Czech Ministry of Education, Youth and Sports from Project No. CZ.02.2.69/0.0/0.0/16\_027/0008371, Czech Ministry of Interior project No. VI20152020025 "DRAPAK", and the National Programme of Sustainability (NPU II) project "IT4Innovations excellence in science - LQ1602". The authors would like to thanks to Mr. Hamid Eghbalzadeh for his valuable discussions.
The authors are also tankful to the DCASE organizers for managing the annual challenges, which allows for rapid advances in the ASC technology.

\bibliographystyle{IEEEtran}
\bibliography{refs}

\end{sloppy}
\end{document}